\documentclass[12pt]{article}
\usepackage{amssymb,amsmath}
\catcode `\@=11
\def\@cite#1#2{#1\if@tempswa , #2\fi}
\catcode `\@=12

\begin{document} 
\title{\Large 
Infinite number of solvable generalizations of XY-chain, \\
with cluster state, 
and with central charge c=m/2
\\
$\:$
} 
\author{Kazuhiko MINAMI}
\date{}
\maketitle
 
\abstract{
An infinite number of spin chains are solved and it is derived 
that the ground-state phase transitions belong to the universality classes 
with central charge $c=m/2$, where $m$ is an integer. 
The models are diagonalized by automatically obtained transformations, 
many of which are different from the Jordan-Wigner transformation. 
The free energies, correlation functions, string order parameters,
exponents, central charges, and the phase diagram are obtained. 
Most of the examples consist of the stabilizers of the cluster state. 
A unified structure of the one-dimensional XY and cluster-type spin chains is revealed, 
and other series of solvable models can be obtained 
through this formula. 
\\
\vspace{3.2cm}

\vspace{0.6cm}

\noindent
Graduate School of Mathematics, Nagoya University, Nagoya, 464-8602, JAPAN

\noindent
minami@math.nagoya-u.ac.jp

\newpage

\section{Introduction}

Lattice spin models are one of the most basic subjects in statistical physics, 
mathematical physics, and condensed matter physics. 
Equivalences or common structures often appear between one-dimensional operators 
and one- and two-dimensional lattice models. 
Onsager solved
[\cite{44Onsager}]
the rectangular Ising model, 
Kaufman and Onsager found 
[\cite{Onsager1949}]
the exact form of the spontaneous magnetization, 
with the help of a one-dimensional operator
[\cite{Onsager1970}] , 
that commutes with the transfer matrix.  
Nambu also solved
[\cite{50Nambu}] 
the rectangular Ising model in 1950;  
a key structure of his derivation was a one-dimensional operator. 
Both operators are a kind of the XY chain.

In the 2000s,  
the cluster state was introduced, 
and investigated 
as a resource state for measurement-based quantum computation (MBQC)
[\cite{01Briegel}][\cite{01Raussendorf}][\cite{03Raussendorf}].
The cluster model is a spin model 
in which the ground state is a cluster state. 
In one-dimension, 
the cluster model was already solved by Suzuki
[\cite{71Suzuki}] 
in 1971, 
and the one-dimensional cluster model (and its generalizations) 
have been investigated by many authors
[\cite{04Pachos}]-[\cite{16Ohta}].
It is known that the free energies 
of the one-dimensional Kitaev model and the one-dimensional cluster model 
are identical
to that of the transverse Ising chain 
(see for example [\cite{07Feng}] and [\cite{09Doherty}][\cite{11Smacchia}]). 

Recently, a simple formula was introduced 
[\cite{16Minami}], 
in which solvable Hamiltonians and transformations to derive the free energy 
can be obtained through the commutation relations. 
The formula is summarized as follows: 
Let us consider a series of operators $\{\eta_j\}$ $(j=1, 2, \ldots, M)$. 
The operators $\eta_j$ and $\eta_k$ are called 'adjacent' 
when $(j, k)=(j, j+1)$ $(1\leq j\leq M-1)$, or $(j, k)=(M, 1)$. 
Let the operators $\eta_j$ satisfy the commutation relations 
\begin{eqnarray}
\eta_{j}\eta_{k}
=
\left\{
\begin{array}{cl}
1 & j=k. \\
-\eta_{k}\eta_{j} & \eta_{j}\: {\rm and}\: \eta_{k} \:{\rm are}\: {\rm adjacent} \\
\eta_{k}\eta_{j} &{\rm otherwise.} \\
\end{array}
\right.
\label{cond}
\end{eqnarray}
Then the Hamiltonian 
\begin{eqnarray}
-\beta{\cal H}
=
\sum_{j=1}^{M}K_j \eta_j
\label{hamKjetaj}
\end{eqnarray}
is mapped to the free fermion system by the transformation 
\begin{eqnarray}
\varphi_j
=
\frac{1}{\sqrt{2}}
e^{i\frac{\pi}{2}(j-1)}
\eta_0
\eta_1
\eta_2
\cdots
\eta_j
\hspace{0.8cm}
(0\leq j\leq M-1), 
\label{transmain}
\end{eqnarray}
where $\eta_0$ is an initial operator satisfying 
$\eta_0^2=-1$,  $\eta_0\eta_1=-\eta_1\eta_0$,  
and $\eta_0\eta_k=\eta_k\eta_0$ $(2\leq k\leq M-1)$, 
which is introduced for convenience. 
We find that $(-2i)\varphi_{j}\varphi_{j+1}=\eta_{j+1}$,
and for all $j$, $k$ that 
\begin{eqnarray}
\{\varphi_j, \varphi_k\}
=
\varphi_j\varphi_k+\varphi_k\varphi_j
=
\delta_{jk}.
\label{anticom}
\end{eqnarray}
Hence the Hamiltonian (\ref{hamKjetaj}) is written 
as a sum of two-body products of the Majorana fermion operators $\varphi_{j}$, 
and thus can be diagonalized. 

The transformation (\ref{transmain}) is automatically generated from $\{\eta_j\}$, 
and only the commutation relation (\ref{cond}) is needed 
to obtain the free energy. 
This procedure is, therefore, quite general. 
The operators 
$\eta_{2j-1}^{(1)}=\sigma_{j}^z$ and $\eta_{2j}^{(1)}=\sigma_{j}^x\sigma_{j+1}^x$ for $j\geq 1$, 
and $M=2N$, $\sigma_{N+1}^x=\sigma_1^x$, $\eta_{0}^{(1)}=i\sigma_{1}^x$, 
satisfy the conditions,  
and in this case (\ref{hamKjetaj}) is the Hamiltonian of the one-dimensional transverse Ising model.   
The operators $\eta_{2j-1}^{(2)}=\sigma_{2j-1}^x\sigma_{2j}^x$ and 
$\eta_{2j}^{(2)}=\sigma_{2j}^y\sigma_{2j+1}^y$ for $j\geq 1$, 
and $M=N$, $\sigma_{N+1}^y=\sigma_1^y$, $\eta_{0}^{(2)}=i\sigma_{1}^y$, 
also satisfy the conditions,  
and in this case (\ref{hamKjetaj}) is the Hamiltonian of the one-dimensional Kitaev model. 
The known fact that the free energies of these two models are identical 
is naturally explained from this formula. 

Hereafter, 
$\varphi_j^{(k)}$ and ${\cal H}^{(k)}$ denote 
$\varphi_j$ and ${\cal H}$ generated from $\{\eta_j^{(k)}\}$, 
respectively, 
where $\{\eta_j^{(k)}\}$ is a series of operators satisfying (\ref{cond}). 
It should be stressed that 
in the cases of $\{\eta_j^{(1)}\}$ and $\{\eta_j^{(2)}\}$, 
the transformation (\ref{transmain}) is intrinsically the Jordan-Wigner transformation. 
However, when we introduce  
$\eta_{2j-1}^{(3)}=\sigma_{2j-1}^x\sigma_{2j}^z\sigma_{2j+1}^x$ and 
$\eta_{2j}^{(3)}=\sigma_{2j}^x 1_{2j+1}\sigma_{2j+2}^x$ $(j\geq 1)$, 
where $1_j$ is the unit operator, 
we obtain a solvable model  generated from this series of operators, 
and together with an initial operator $\eta_{0}^{(3)}=i\sigma_{1}^x\sigma_{2}^x$, 
we find that the transformation (\ref{transmain}) 
that diagonalize (\ref{hamKjetaj}) is 
\begin{eqnarray}
\varphi^{(3)}_{2j}
&=&
\frac{1}{\sqrt{2}}
(\prod_{\nu=1}^{j}1_{2\nu-1}\sigma_{2\nu}^z)
\sigma_{2j+1}^x\sigma_{2j+2}^x,
\nonumber
\\
\varphi^{(3)}_{2j+1}
&=&
\frac{1}{\sqrt{2}}
(\prod_{\nu=1}^{j}1_{2\nu-1}\sigma_{2\nu}^z)
1_{2j+1}\sigma_{2j+2}^y\sigma_{2j+3}^x
\hspace{1.0cm}
(j=0, 1, 2, 3, \ldots).
\label{trans3}
\end{eqnarray}
When we introduce  
$\eta_{2j-1}^{(4)}=\sigma_{4j-3}^x\sigma_{4j-2}^z\sigma_{4j-1}^x$ and 
$\eta_{2j}^{(4)}=\sigma_{4j-1}^z 1_{4j}\sigma_{4j+1}^z$ $(j\geq 1)$, 
we obtain another solvable model, 
and together with an initial operator $\eta_{0}^{(4)}=(-i)\sigma_{1}^z$, 
we find that the transformation (\ref{transmain}) 
that diagonalize (\ref{hamKjetaj}) is 
\begin{eqnarray}
\varphi^{(4)}_{2j}
&=&
\frac{(-1)^{j-1}}{\sqrt{2}}
(\prod_{\nu=1}^{j} \sigma_{4\nu-3}^y \sigma_{4\nu-2}^z \sigma_{4\nu-1}^y 1_{4\nu} )
\sigma_{4j+1}^z,
\nonumber\\
\varphi^{(4)}_{2j+1}
&=&
\frac{(-1)^{j}}{\sqrt{2}}
(\prod_{\nu=1}^{j} \sigma_{4\nu-3}^y \sigma_{4\nu-2}^z \sigma_{4\nu-1}^y 1_{4\nu} )
\sigma_{4j+1}^y \sigma_{4j+2}^z \sigma_{4j+3}^x
\nonumber
\\
&&
\hspace{7.0cm}
(j=0, 1, 2, 3, \ldots).
\label{trans4}
\end{eqnarray}
Transformations (\ref{trans3}) and (\ref{trans4}) 
are different from the Jordan-Wigner transformation. 
These transformations are automatically generated from the series of operators.

The one-dimensional XY model,  the two-dimensional Ising model,  
and other unsolved composite models 
can be diagonalized through this formula[\cite{16Minami}].
The Jordan-Wigner transformation is a special case of (\ref{transmain}). 
In this paper, 
the diagonalization method in [\cite{16Minami}] is applied to the XY and cluster type interactions, 
and an infinite number of solvable spin chains are obtained. 
Specifically, extreme generalizations of the XY chain are investigated, 
their correlation functions and critical exponents are obtained, 
and it is derived that these models show a universal phase diagram 
and common critical behavior with central charge $c=m/2$. 
The arguments written in terms of $\eta_j$ and $\varphi_j$ 
are valid for all of them, 
and therefore an unified structure of an infinite number of spin chains is found through this formula. 
\\

\section{Model and the free energy}
Let $N$ be the number of sites, 
and $N/2$ and $M/4$ be positive integers. 
Let us introduce the notation  
$\varphi_{2j-2}=\varphi_1(j)$ and $\varphi_{2j-1}=\varphi_2(j)$. 
In the case of $\{\eta_j^{(1)}\}$, 
we find  
$(+2i)\varphi^{(1)}_2(j)\varphi^{(1)}_1(j)=\sigma_{j}^z$,  
$(-2i)\varphi^{(1)}_2(j)\varphi^{(1)}_1(j+1)=\sigma_{j}^x\sigma_{j+1}^x$. 
Let us then generally consider the Hamiltonian 
which is composed by two-body products of $\varphi_\tau(j)$ as 
\begin{eqnarray}
-\beta{\cal H}
&=&
\sum_l K_l \sum_{j=1}^{M/2} (-2i)\varphi_2(j)\varphi_1(j+l)
\nonumber\\
&+&
\sum_{l>0} K_l^{(1)}\sum_{j=1}^{M/2} (+2i)\varphi_1(j)\varphi_1(j+l)
\nonumber\\
&+&
\sum_{l>0} K_l^{(2)} \sum_{j=1}^{M/2} (-2i)\varphi_2(j)\varphi_2(j+l).
\label{Hamiltonian}
\end{eqnarray}
Each term in (\ref{Hamiltonian}) is written as a product of $\eta_j$, for example, 
\begin{eqnarray}
(-2i)\varphi_2(j)\varphi_1(j+1)
=
\eta_{2j},
\nonumber\\
(+2i)\varphi_2(j)\varphi_1(j)
=
\eta_{2j-1},
\end{eqnarray}
and generally 
\begin{eqnarray}
&&
(-2i)\varphi_2(j)\varphi_1(j+l)
=
(-1)^{l-1}\eta_{2j}\eta_{2j+1}\cdots\eta_{2j+2l-2},
\nonumber\\
&&
(-2i)\varphi_2(j)\varphi_1(j-l)
=
(-1)^{l-1}\eta_{2j-2l-1}\eta_{2j-2l}\cdots\eta_{2j-1}.
\end{eqnarray}
In the case of the operators $\{\eta_j^{(1)}\}$, we find for example 
\begin{eqnarray}
(-2i)\varphi^{(1)}_2(j)\varphi^{(1)}_1(j+3)
&=&
\sigma^x_{j}\sigma^z_{j+1}\sigma^z_{j+2}\sigma^x_{j+3}
\nonumber
\\
(+2i)\varphi^{(1)}_2(j)\varphi^{(1)}_1(j+2)
&=&
\sigma^x_{j}\sigma^z_{j+1}\sigma^x_{j+2}
\nonumber
\\
(-2i)\varphi^{(1)}_2(j)\varphi^{(1)}_1(j+1)
&=&
\sigma^x_{j}\sigma^x_{j+1}
\nonumber
\\
(+2i)\varphi^{(1)}_2(j)\varphi^{(1)}_1(j)
&=&
\sigma^z_{j}
\nonumber
\\
(-2i)\varphi^{(1)}_2(j)\varphi^{(1)}_1(j-1)
&=&
\sigma^y_{j-1}\sigma^y_{j}
\nonumber
\\
(+2i)\varphi^{(1)}_2(j)\varphi^{(1)}_1(j-2)
&=&
\sigma^y_{j-2}\sigma^z_{j-1}\sigma^y_{j}
\nonumber
\\
(-2i)\varphi^{(1)}_2(j)\varphi^{(1)}_1(j-3)
&=&
\sigma^y_{j-3}\sigma^z_{j-2}\sigma^z_{j-1}\sigma^y_{j}
\label{series1}
\end{eqnarray}
Thus the transverse Ising model, the XY model, and the cluster models are generated 
from $\{\eta_j^{(1)}\}$. 
The interactions (\ref{series1}) are those introduced and diagonalized in [\cite{71Suzuki}]. 
Each series of operators $\{\eta^{(k)}_j\}$ that satisfy (\ref{cond}) 
thus provides an infinite number of solvable Hamiltonians; 
other series of new solvable interactions can be generated from other $\{\eta^{(k)}_j\}$.

There appear, however, an additional factor at the boundary, for example,   
\begin{eqnarray}
(-2i)\varphi_2(M/2)\varphi_1(1)=(i^M\eta_1\cdots\eta_M)\eta_M. 
\end{eqnarray}
It is straightforward to convince that 
the cyclic boundary condition for $\eta_j$ yields the definition that 
\begin{eqnarray}
\varphi_\tau(M/2+l)
=
\left\{
\begin{array}{cl}
+\varphi_\tau(l) & \hspace{0.3cm}i^M\eta_1\cdots\eta_M=+1\\
-\varphi_\tau(l) & \hspace{0.3cm}i^M\eta_1\cdots\eta_M=-1,
\end{array}
\right.
\label{bcond}
\end{eqnarray}
where $\tau=1, 2$ 
and $i^M\eta_1\cdots\eta_M$ is hermitian and its eigenvalues are $\pm 1$. 

Let us introduce Fourier transformations of $\varphi_\tau(j)$ as 
\begin{eqnarray}
\varphi_\tau(j)
=
\frac{1}{\sqrt{M/2}}\sum_{0<q<\pi}(e^{iqj}C_\tau(q)+e^{-iqj}C_\tau^\dag(q)),
\label{fouriertrans}
\end{eqnarray}
where 
\begin{eqnarray}
\{C_r^\dag(p), C_s(q)\}
=
\delta_{pq}\delta_{rs},
\hspace{0.4cm}
\{C_r(p), C_s(q)\}=0.
\label{comcpcq}
\end{eqnarray}
According to (\ref{bcond}), $\displaystyle q=\frac{\rho}{M/2}\pi$, 
where $\rho=$even and $\rho=$odd 
for $i^M\eta_1\cdots\eta_M$ is $+1$ and $-1$, respectively. 
The Hamiltonian (\ref{Hamiltonian}) is expressed as 
\begin{eqnarray}
-\beta{\cal H}
&=&
\sum_{0<q<\pi}
(+2i)[L(q)C_1(q)C_2^\dag(q)+L(q)^\dag C_1^\dag(q)C_2(q)],
\nonumber\\
&+&
\sum_{0<q<\pi}
(+2i)L_1(q)(C_1^\dag(q)C_1(q)-\frac{1}{2})
\nonumber\\
&+&
\sum_{0<q<\pi}
(-2i)L_2(q)(C_2^\dag(q)C_2(q)-\frac{1}{2}),
\label{HamCdagC}
\end{eqnarray}
where 
\begin{eqnarray}
L(q)
&=&
\sum_l K_l e^{iql}, 
\nonumber\\
L_\tau(q)
&=&
\sum_{l>0} K_l^{(\tau)} (e^{iql}-e^{-iql}) 
\hspace{1.4cm}
(\tau=1,2),
\end{eqnarray}
and $L(q)^\dag$ is the complex conjugate of $L(q)$. 
We have used the relation $\sum_{0<q<\pi}\cos ql=0$. 
Hamiltonian (\ref{HamCdagC}) is written as a sum of contributions from each $q$, 
which commute with each other, 
and which can be expressed as a block-diagonal matrix 
with respect to  the bases 
$|0\rangle$, $C_1^\dag(q)|0\rangle$, $C_2^\dag(q)|0\rangle$, $C_2^\dag(q)C_1^\dag(q)|0\rangle$, 
as
\begin{eqnarray}
\sum_{0<q<\pi}
\left[
\begin{array}{cccc}
-iL_1(q)+iL_2(q) & 0 &0 & 0 \\
0 & iL_1(q)+iL_2(q)  & 2iL(q)^\dag  &0 \\
0 & -2iL(q) &-iL_1(q)-iL_2(q)  & 0  \\
0 & 0 &0 & iL_1(q)-iL_2(q) 
\end{array}
\right]_{q.}
\label{matrix}
\end{eqnarray}
Note that $L_1(q)$ and $L_2(q)$ are pure imaginary when $K_l^{(1)}$ and $K_l^{(2)}$ are real. 
Diagonalization of (\ref{matrix}) corresponds to the Bogoliubov transformation. 
The spin space is a $2^N$-dimensional Hilbert space, 
and here we find $M/4$ different $q$ in $0<q<\pi$, 
and therefore $4^{M/4}$ eigenvalues. 
Then we find $g_N=2^N/4^{M/4}=2^{N-M/2}$ -fold degeneracy for each state. 
The partition function is obtained from the eigenvalues as 
\begin{eqnarray}
Z
=
g_N
\prod_{0<q<\pi}(e^{\lambda(q)}+e^{-\lambda(q)}+e^{\Lambda(q)}+e^{-\Lambda(q)}),
\end{eqnarray}
where 
\begin{eqnarray}
\lambda(q)
&=&
i L_1(q)-iL_2(q),
\nonumber\\
\Lambda(q)
&=&
[4LL^\dag+(i L_1(q)+iL_2(q))^2]^{1/2}.
\end{eqnarray}

Let us, for example, consider $\{\eta_j^{(1)}\}$, 
where $M=2N$. 
Let $K_0=-h$, $K_1=K$, and the other coefficients be zero in (\ref{Hamiltonian}), 
then we obtain     
\begin{eqnarray}
-\beta f
&=&
\lim_{N\to\infty}\frac{\log Z}{N}
=
\frac{1}{\pi}\int_0^{\pi} 
\log(e^{\frac{1}{2}\Lambda(q)}+e^{-\frac{1}{2}\Lambda(q)})\:dq, 
\nonumber\\
\Lambda(q)
&=&
2\sqrt{K^2+h^2-2hK\cos q}, 
\label{freeengtrIsing}
\end{eqnarray}
which is the free energy of the transverse Ising chain. 
When $K_\kappa=-h$, $K_{\kappa+m}=K$ $(m\neq 0)$, and the other coefficients are all zero, 
it is easy to show that 
the free energy is identical to (\ref{freeengtrIsing}). 
(Generally if $L_1(q)=L_2(q)=0$, 
the free energy is invariant under $L(q)\mapsto e^{i\kappa q}L(q)$.)  
The Hamiltonian of the transverse Ising chain, 
and the one-dimensional cluster models investigated in [\cite{09Doherty}] and [\cite{11Smacchia}], 
are generated from $\{\eta_j^{(1)}\}$, 
with $(\kappa, m)=(0,1)$, $(0,2)$ and $(-1,3)$, respectively. 
The coincidences of their free energies are explained from this invariance.

We already noticed from (\ref{series1}) that the XY model is generated from $\{\eta_j^{(1)}\}$. 
Let us, however, consider $\{\eta_j^{(2)}\}$ 
and introduce the shifted operators 
$\bar{\eta}_{2j-1}^{(2)}=\sigma_{2j-1}^y\sigma_{2j}^y$ and 
$\bar{\eta}_{2j}^{(2)}=\sigma_{2j}^x\sigma_{2j+1}^x$. 
Then 
${\cal H}^{(2)}+\bar{{\cal H}}^{(2)}$ 
involves the interactions 
$K_x(+2i)\varphi_2^{(2)}(j)\varphi_1^{(2)}(j)+K_y(+2i)\bar{\varphi}_2^{(2)}(j)\bar{\varphi}_1^{(2)}(j)=K_x\sigma_{2j-1}^x\sigma_{2j}^x+K_y\sigma_{2j-1}^y\sigma_{2j}^y$, 
and thus complete XY interactions are realized from two Kitaev chains. 
The XY model, therefore, is also generated from 
$\{\eta_j^{(2)}\}$ and $\{{\bar\eta}_j^{(2)}\}$. 
Operators $\eta_j^{(2)}$ and $\bar{\eta}_k^{(2)}$ commute for all $j$, $k$, 
and hence 
${\cal H}^{(2)}+\bar{{\cal H}}^{(2)}$
can be diagonalized with our formula.  

Let us here consider Table 1 and Table 2. 
In the case of  $(k, n, l)=(3, 1, 1)$ in Table 2, for example, 
we find the operators  
$\eta_{2j-1}^{(3, 1, 1)}=\sigma_{2j-1}^x\sigma_{2j}^z\sigma_{2j+1}^x$ and 
$\eta_{2j}^{(3, 1, 1)}=\sigma_{2j}^x 1_{2j+1}\sigma_{2j+2}^x$.  
In this case, we can also introduce the shifted operators 
$\bar{\eta}_{2j-1}^{(3, 1, 1)}=\sigma_{2j}^x\sigma_{2j+1}^z\sigma_{2j+2}^x$ and 
$\bar{\eta}_{2j}^{(3, 1, 1)}=\sigma_{2j+1}^x 1_{2j+2}\sigma_{2j+3}^x$.  
Operators $\eta_j^{(3, 1, 1)}$ and $\bar{\eta}_k^{(3, 1, 1)}$ commute for all $j$, $k$, 
and two series of operators $\{\eta_j^{(3, 1, 1)}\}$ and $\{{\bar\eta}_j^{(3, 1, 1)}\}$ 
generate the complete Hamiltonian $(k, n, l)=(3, 1, 1)$ in Table 1. 
Several series of shifted operators are generally needed to construct total Hamiltonian. 

Examples of other solvable Hamiltonians are listed in Table 1. 
Some examples of $\{\eta^{(k, n, l)}_j\}$, 
and other interactions that can be found in (\ref{Hamiltonian}), 
are listed in Table 2. 
The operators 
$\eta^{(1, 0, 1)}_j$, $\eta^{(2, 1, 0)}_j$, $\eta^{(3, 1, 1)}_j$, and $\eta^{(4, 1, -)}_j$ 
are those written as 
$\eta^{(1)}_j$, $\eta^{(2)}_j$, $\eta^{(3)}_j$, and $\eta^{(4)}_j$, respectively,  
in Introduction. 
The initial operators $\eta^{(k, n, l)}_0$ 
and the transformations $\varphi^{(k, n, l)}_{j}$ that diagonalize (\ref{Hamiltonian}) 
are also given in Table 2. 
The models except the cases $(k, n, l)=(1, 0, 1)$ and $(2, 1, 0)$ 
cannot be solved by the Jordan-Wigner transformation. 
\\

\section{Ground state correlation functions}
Operators $(-2i)\varphi_2(j)\varphi_1(j+\kappa)$ in the Hamiltonian (\ref{Hamiltonian}) 
commute with each other and form stabilizers[\cite{96Gottesman}][\cite{97Calderbank}], 
i.e. 
\begin{eqnarray}
(-2i)\varphi_2(j)\varphi_1(j+\kappa)|\Phi\rangle
=
\epsilon_j |\Phi\rangle,
\end{eqnarray}
where $\epsilon_j=+1$ or $-1$, 
and a set of values  $\{\epsilon_j\}$ specify a subspace of the Hilbert space. 
Specifically, 
we find $(-2i)\varphi^{(1)}_2(j)\varphi^{(1)}_1(j+2)=\sigma_{j}^x\sigma_{j+1}^z\sigma_{j+2}^x$, 
and the state $|\Phi\rangle$ 
that satisfy 
$\sigma_{j}^x\sigma_{j+1}^z\sigma_{j+2}^x|\Phi\rangle=|\Phi\rangle$ 
for all $j$ 
is the cluster state. 
(Operators 
$\{(+2i)\varphi^{(3)}_2(j)\varphi^{(3)}_1(j)\}$ 
together with the shifted operators 
$\{(+2i){\bar \varphi}^{(3)}_2(j){\bar \varphi}^{(3)}_1(j)\}$ 
also form stabilizers of the cluster state.)  

The string order parameters are obtained from products of 
$\{(-2i)\varphi_2(j)\varphi_1(j+\kappa)\}$: 
\begin{eqnarray}
I_\kappa(n)
=
\langle \prod_{j=i}^{i+n-1}(-2i)\varphi_2(j)\varphi_1(j+\kappa) \rangle_0,
\label{StringCorr}
\end{eqnarray}
where $\langle\:\rangle_0$ is the ground-state expectation value. 
Non-vanishing value of $I_\kappa(n)$ indicates 
the presence of the order stabilized by these operators. 
In the case of $\{\eta_j^{(1)}\}$, for example,  
\begin{eqnarray}
&&
I^{(1)}_1(n)
=
\langle \prod_{j=i}^{i+n-1}\sigma_{j}^x\sigma_{j+1}^x \rangle_0
=
\langle \sigma_{i}^x\sigma_{i+n}^x \rangle_0,
\nonumber\\
&&
I^{(1)}_2(n)
=
\langle \prod_{j=i}^{i+n-1}\sigma_{j}^x\sigma_{j+1}^z \sigma_{j+2}^x \rangle_0
\nonumber\\
&&
=
(-1)^{n-2}
\langle \sigma_{i}^x\sigma_{i+1}^y(\sigma_{i+2}^z\cdots\sigma_{i+n-1}^z)\sigma_{i+n}^y\sigma_{i+n+1}^x \rangle_0.
\label{StringCorr1}
\end{eqnarray}
In the case of $\{\eta_j^{(2)}\}$, 
\begin{eqnarray}
&&
I^{(2)}_1(n)
=
\langle 
\prod_{j=i}^{i+n-1}\sigma_{2j}^y\sigma_{2j+1}^y 
\rangle_0
=
\langle
\sigma_{2i}^y\sigma_{2i+1}^y\cdots\sigma_{2i+2n-1}^y
\rangle_0,
\nonumber\\
&&
I^{(2)}_2(n)
=
\langle 
\prod_{j=i}^{i+n-1}(-1)\sigma_{2j}^y\sigma_{2j+1}^z\sigma_{2j+2}^z\sigma_{2j+3}^y 
\rangle_0
\nonumber\\
&&
=
(-1)^{n}
\langle
\sigma_{2i}^y\sigma_{2i+1}^z
(\sigma_{2i+2}^x\cdots\sigma_{2i+2n-1}^x)
\sigma_{2i+2n}^z\sigma_{2i+2n+1}^y 
\rangle_0.
\end{eqnarray}
Here $I^{(k)}_\kappa(n)$ is a correlation function of the model 
generated from $\{\eta^{(k)}_j\}$. 
Note that $I^{(k)}_\kappa(n)$ is independent of $k$, 
because there remains no difference when they are written by $\eta_j$. 
Stabilizers and the corresponding string order parameters are listed in Table 3.

Let us consider the case $L_1(q)=L_2(q)=0$. 
The Hamiltonian (\ref{HamCdagC}) is diagonalized 
by the canonical transformation  
\begin{eqnarray}
\xi_1(q)
&=&
(iL C_1(q)-\sqrt{LL^\dag}C_2(q))/\sqrt{2LL^\dag},
\nonumber\\
\xi_2(q)
&=&
(iL C_1(q)+\sqrt{LL^\dag}C_2(q))/\sqrt{2LL^\dag}, 
\end{eqnarray}
as 
\begin{eqnarray}
-\beta{\cal H}
&=&
2\sqrt{LL^\dag}
\sum_{0<q<\pi}
(\xi_1(q)^\dag\xi_1(q)-\xi_2(q)^\dag\xi_2(q)).
\end{eqnarray}
The ground state is realized 
when $\xi_1(q)^\dag\xi_1(q)=1$, $\xi_2(q)^\dag\xi_2(q)=0$ for all $q$, 
and in this case 
$\langle\xi_1(q)^\dag\xi_1(q)\rangle_0=1$, 
$\langle\xi_1(q)^\dag\xi_2(q)\rangle_0=\langle\xi_1(q)\xi_2(q)^\dag\rangle_0=\langle\xi_2(q)^\dag\xi_2(q)\rangle_0=0$. 
The ground state correlation function is obtained as 
\begin{eqnarray}
&&
\frac{1}{M/2}
\langle \sum_{j=1}^{M/2} (-2i)\varphi_2(j)\varphi_1(j+\kappa) \rangle_0
\nonumber\\
&&
=
(+2i)
\frac{1}{M/2}
\sum_{0<q<\pi}
(e^{iq\kappa}\langle C_1(q)C_2(q)^\dag \rangle_0+e^{-iq\kappa}\langle C_1(q)^\dag C_2(q) \rangle_0)
\nonumber\\
&&
=
\frac{1}{M/2}
\sum_{0<q<\pi}
(e^{iq\kappa}\frac{\sqrt{LL^\dag}}{L}+e^{-iq\kappa}\frac{\sqrt{LL^\dag}}{L^\dag})
(\langle \xi_1(q)^\dag\xi_1(q)\rangle_0-\langle \xi_2(q)^\dag\xi_2(q)\rangle_0)
\nonumber\\
&&
=
\frac{1}{M/2}
\sum_{0<q<\pi}
\frac{L^\dag e^{iq\kappa}+L e^{-iq\kappa}}{\sqrt{LL^\dag}}
\end{eqnarray}
In the limit $M\rightarrow\infty$, 
\begin{eqnarray}
\langle (-2i)\varphi_2(j)\varphi_1(j+\kappa) \rangle_0
=
\frac{1}{2\pi}
\int_0^\pi 
\frac{L^\dag e^{iq\kappa}+L e^{-iq\kappa}}{\sqrt{LL^\dag}}
\: dq.
\label{Corrfn}
\end{eqnarray}
One can also derive that 
\begin{eqnarray}
\langle (-2i)\varphi_1(j)\varphi_1(j+\kappa) \rangle_0
=
\langle (-2i)\varphi_2(j)\varphi_2(j+\kappa) \rangle_0
=0
\hspace{0.6cm}
(\kappa\neq 0)
\label{Corrfn1122}
\end{eqnarray}
at the ground state.

Because of the expression of $\varphi_\tau(j)$ given in (\ref{fouriertrans})-(\ref{comcpcq}), 
and the fact (\ref{Corrfn1122}), 
and with the use of the Wick's theorem, 
the string order parameter (\ref{StringCorr}) is expressed as 
\begin{eqnarray}
\sum_P ({\rm sgn}\: P) 
\prod_{j=i}^{i+n-1}\langle (-2i)\varphi_2(j)\varphi_1(P(j+\kappa)) \rangle_0,
\label{detG}
\end{eqnarray}
where $P$ are the permutations of the indices. 
The sum (\ref{detG}) is the determinant of the matrix $G_n$, 
where $(G_n)_{\mu\nu}=\langle (-2i)\varphi_2(j+\mu-1)\varphi_1(j+\kappa+\nu-1) \rangle_0$. 
\\


\section{Ground state phase transitions}
Let us here consider the Hamiltonian 
\begin{eqnarray}
-\beta{\cal H}
&=&
K_\kappa\sum_{j=1}^{M/2}
\varphi_2(j)\varphi_1(j+\kappa)
\label{HamgenXY}
\\
&+&
K_{\kappa-m}\sum_{j=1}^{M/2}
\varphi_2(j)\varphi_1(j+\kappa-m)
+
K_{\kappa+m}\sum_{j=1}^{M/2}
\varphi_2(j)\varphi_1(j+\kappa+m),
\nonumber
\end{eqnarray}
where $m$ is a positive integer. 
This Hamiltonian (\ref{HamgenXY}) is a generalization of the XY chain, 
and includes many important models. 
Let $(k, n, l)=(1, 0, 1)$ in Table 2. 
Then the case $\kappa=0$ and $m=1$ is the XY chain with an external field. 
The case $\kappa=1$ and $m=1$ includes the cluster model with the Ising interaction. 
Let $(k, n, l)=(2, 1, 0)$, then the case $\kappa=0$ and $m=1$ 
is the one-dimensional Kitaev model with an additional interaction. 
Let $(k, n, l)=(3, 1, 1)$, then the case $\kappa=0$ and $m=1$ 
includes the cluster model with next-nearest neighbor interaction. 
The cases $(k, n, l)=(1, 0, 1)$, $(3, 1, 1)$, $(3, 1, 2)$, $(3, 1, 3)$, $(4, 1, -)$, and $(11, -, -)$ 
include the stabilizers of the cluster state, 
and various generalizations of the one-dimensional cluster model 
are obtained in the framework of (\ref{HamgenXY}). 

In the case of (\ref{HamgenXY}), we find 
\begin{eqnarray}
L(q)
=
K_{\kappa-m}e^{iq(\kappa-m)}+K_\kappa e^{iq\kappa}+K_{\kappa+m}e^{iq(\kappa+m)}.
\end{eqnarray}
Let 
$K_{\kappa}=-K\neq 0$,  
$K_{\kappa+m}=K(1+\gamma_1)/2$,  
$K_{\kappa-m}=K(1-\gamma_2)/2$, 
and consider the asymptotic form of $I_\kappa(n)$. 
Let $\displaystyle (G_n)_{\mu\nu}=\frac{K_\kappa}{|K_\kappa|}{(M_n)_{\mu\nu}}$. 
The matrix $M_n$ is a Toeplitz determinant, 
i.e. the $(\mu, \nu)$ element depends only on 
$\nu-\mu=\tau$, $(M_n)_{\mu\nu}=(M_n)_\tau$, 
and thus the Szeg\H{o}'s theorem can be applied. 
Let us introduce $f(p)$ by the relation 
\begin{eqnarray}
(M_n)_\tau
=
\frac{1}{2\pi}\int_{-\pi}^{\pi} e^{-ip\tau} f(p)\:dp. 
\label{Mntau}
\end{eqnarray}
Then the Szeg\H{o}'s theorem says that 
the asymptotic form of $\det M_n$ is   
\begin{eqnarray}
\lim_{n\to\infty}\frac{\det M_n}{\lambda^n}
=
\exp(\sum_{n=1}^\infty ng_{n}g_{-n}),
\end{eqnarray}
where 
\begin{eqnarray}
g_n
=
\frac{1}{2\pi}\int_{-\pi}^{\pi} e^{-ipn} \ln f(p)\:dp,
\end{eqnarray}
and $\lambda=\exp g_0$. 
We obtain, from (\ref{Corrfn}) and  (\ref{Mntau}), that 
\begin{eqnarray}
&&
f(p)
=
\frac{L(p)}{\sqrt{L(p)L^\dag(p)}}e^{-ip\kappa}
\Big/ \frac{K_\kappa}{|K_\kappa|}
\\
&&
=
\frac{1+\alpha_1 e^{ipm}+\alpha_2 e^{-ipm}}
{[(1+\alpha_1 e^{-ipm}+\alpha_2 e^{ipm})(1+\alpha_1 e^{ipm}+\alpha_2 e^{-ipm})]^{1/2}},
\nonumber
\end{eqnarray}
where 
$\alpha_1=K_{\kappa+m}/K_\kappa$ and $\alpha_2=K_{\kappa-m}/K_\kappa$. 
With the use of the formula 
\begin{eqnarray}
1+\alpha_1 e^{ipm}+\alpha_2 e^{-ipm}
=
A(1+a_1 e^{ipm})(1+a_2 e^{-ipm}),
\end{eqnarray}
where 
$a_1=\alpha_1t$, $a_2=\alpha_2t$, 
$t=2/(1+\sqrt{1-4\alpha_1\alpha_2})$, 
the factor $A$ is independent of $p$, 
and the formula 
\begin{eqnarray}
&&
\frac{1}{2\pi}\int_{-\pi}^{\pi} e^{-ipn} \ln (1+a\: e^{ipm})\:dp
=
\frac{1}{2\pi i}\int_{C} \sum_{l=1}^\infty  \frac{(-1)^{l-1}a^l}{l\: z^{1+n-ml} }\:dz
\nonumber\\
&&
\hspace{1.6cm}
=
\left\{
\begin{array}{cl}
(-1)^{l-1}a^l/l & n=ml,\: l\geq 1 \\
0 & {\rm otherwise,} 
\end{array}
\right.
\end{eqnarray}
where $|a|<1$, $z=e^{ip}$, and $C$ is the unit circle, 
we obtain, for $|a_1|<1$ and $|a_2|<1$, that 
\begin{eqnarray}
g_n
=
\left\{
\begin{array}{cl}
(-a_2)^l/2l-(-a_1)^l/2l& n=ml>0\\
0 & n\neq ml>0,
\end{array}
\right.
\end{eqnarray}
$g_{-n}=-g_n$ $(n\neq 0)$, and $g_0=0$.  
Thus we obtain 
\begin{eqnarray}
I_\kappa(n)
\simeq
(\frac{K_\kappa}{|K_\kappa|})^n [(1-a_1^2)(1-a_2^2)(1-a_1a_2)^{-2}]^{m/4}.
\label{I_infty}
\end{eqnarray}
If $|a_1|>1$ or $|a_2|>1$, 
it is easy to show that $\sum_n ng_ng_{-n}\to-\infty$ and $I_\kappa(n)\to 0$. 
Similarly, we can derive the asymptotic form of $I_{\kappa\pm m}(n)$, 
and the results are summarized in Figure 1. 

Specifically in the case of $\{\eta_j^{(1)}\}$, $\kappa=1$, $m=1$, and $K_2=0$ 
yield the transverse Ising chain, 
and from (\ref{StringCorr1}), 
\begin{eqnarray}
\langle\sigma^x_{i}\rangle_0^2
=
\lim_{n\to\infty}|\langle\sigma^x_{i}\sigma^x_{i+n}\rangle_0|
\simeq
(1-(K_0/K_1)^2)^{1/4}.
\end{eqnarray}
We thus obtain $\beta=1/8$, 
which is also the critical exponent of the spontaneous magnetization 
of the rectangular Ising model.

\section{Central charge}
Let us consider the central charge  
of the transition on the critical line $\gamma_2=\gamma_1\neq 0$. 
Let $M=\tau N$ and write $\gamma_1=\gamma$. 
From the dispersion relation 
\begin{eqnarray}
\Lambda(q)
=
2\sqrt{LL^\dag}
\simeq
2|\gamma K|m\frac{2}{\tau} |\frac{l}{N}\pi|
\hspace{0.6cm}
(\frac{l}{N}\pi=p\simeq 0)
\end{eqnarray}
the conformal invariant normalization of the Hamiltonian is obtained 
[\cite{86GehlenRittenbergRuegg}]
from the condition
\begin{eqnarray}
4|\gamma K|m/\tau =1.
\label{confnorm}
\end{eqnarray}
The central charge can be obtained 
[\cite{86BloteCardyNightingale}]
[\cite{86Affleck}] 
(see also [\cite{87Henkel}]) 
from the finite size effect of the ground-state energy: 
\begin{eqnarray}
E_0
=-\sum_{0<q<2\pi}2|\gamma K||\sin\frac{mq}{2}|
[1-\frac{\gamma^2-1}{\gamma^2}\sin^2\frac{mq}{2}]^{1/2}.
\label{E0N}
\end{eqnarray}
With the use of (\ref{confnorm}), and the relations such as 
\begin{eqnarray}
\sum_{k=1}^{M/2}
|\sin\frac{mq}{2}|
=
m\:{\rm Im}
\sum_{k=1}^{M/2m} e^{i\frac{(2k-1)\pi}{M/m}}
=
\frac{m}{\sin\frac{m\pi}{M}},
\end{eqnarray}
where $q=\frac{(2k-1)\pi}{M/2}$, 
the term proportional to $1/N$ in (\ref{E0N}) is obtained as
\begin{eqnarray}
-2|\gamma K| m\:
\frac{1}{6}\frac{m\pi}{\tau N}
=
-\frac{m\pi}{12N}.
\end{eqnarray}
This term should be equal to $-c\pi/6N$, 
which results in the central charge $c=m/2$. 
Similarly on the critical line $\gamma_2=-\gamma_1< 0$, 
we obtain $c=2m/2$. 
The XY chain
[\cite{87Henkel}] 
and the cluster models investigated in 
[\cite{11Smacchia}] 
and
[\cite{15Lahtinen}] 
are obtained from $\{\eta_j^{(1)}\}$, 
and correspond to the cases with 
$m=1$,  $m=3$, and general $m$, 
respectively.

One can convince that the operators in (\ref{HamgenXY}) are classified into $m$ number of subsets 
$\{ \varphi_2(\rho+ml)\varphi_1(\rho+ml), \varphi_2(\rho+ml)\varphi_1(\rho+ml\pm m),  l=1, 2, 3,\ldots\}$, 
where $\rho=1, 2, \ldots, m$. 
The operators with different $\rho$ commute with each other, 
and thus the system is decoupled into $m$ number of XY chains,  
though all the $\{\varphi_2(j)\varphi_1(j)\}$, for example, are needed 
to form complete stablizers of its ground state. 

It is known that 
the models with the interactions (\ref{series1}), 
which are generated from $\{\eta^{(1)}_j\}$, 
can be regarded as coupled transverse Ising chains[\cite{15Lahtinen}]. 
The universality classes of these cases are found in [\cite{15Lahtinen}]  
(see also [\cite{13Mansson}][\cite{14Lahtinen}]). 
The fact that the local Hamiltonians in (\ref{HamgenXY}) are classified into $m$ number of sub-series  
corresponds to this structure. 
The derivation in [\cite{15Lahtinen}] 
relies on the explicit expression of $\{\eta^{(1)}_j\}$ by the Pauli operators, 
and valid for the series generated from $\{\eta^{(1)}_j\}$. 
Here it is derived the same decouplings are found in the 
models generated from other $\{\eta^{(k)}_j\}$, 
and correspondingly we obtain the central charge $c=m/2$. 

\section{Conclusion}
In this paper, 
we considered and applied a general framework 
by which series of solvable spin models can be obtained. 
We construct models 
which consist of stabilizers, and thus have corresponding ground states, 
for example the cluster state. 
The free energies, correlation functions, string order parameters,
exponents, central charges, and the phase diagram are derived exactly. 
This formula clarifies a systematic structure of one-dimensional generalized cluster-type models,  
which are also extreme generalizations of the XY chain. 
We show, in Table 1, examples of solvable Hamiltonians,  
and we show, in Table 2,  examples of operators 
which satisfy (\ref{cond}), 
each of which generates infinite series of solvable spin chains, 
and many of which involve stabilizers of the cluster state. 
It is easy to find series of operators that satisfy (\ref{cond}), 
and thus one can obtain an infinite number of solvable Hamiltonians, 
with ground-state phase transitions with central charge $c=m/2$.

It is known that 
the one-dimensional cluster state is a state 
in which 
the localizable entanglement length  
[\cite{04Verstraete}][\cite{05PoppVerstraete}]
is infinite
[\cite{09Skrovseth}], 
and hence considered to be a candidate for quantum computational wire
[\cite{10Gross}].
In Table 2, the operators $\sigma^x_{j}\sigma^z_{j+1}\sigma^x_{j+2}$ are stabilizers of the cluster state. 
These models may serve as candidates for quantum devices
[\cite{12Else}].

This formula basically relies only on the commutation relations, 
and hence $\{\eta_j\}$ do not need to be spin operators, 
is applicable to any space dimensions, 
and ${\cal H}$ does not need to be an operator related to equilibrium problems. 
The transformation (\ref{transmain}) 
is a stepping stone 
to go beyond the Jordan-Wigner transformation. 

The author would like to thank Prof.~M.~Suzuki for his valuable comment. 
\\




\begin{thebibliography}{00}


\bibitem{44Onsager}
L. Onsager, 
Phys. Rev. 65, 117 (1944).

\bibitem{Onsager1949}
L. Onsager, 
Nuovo Cimento (Suppl.) 6, 261 (1949).

\bibitem{Onsager1970}
L. Onsager, 
in "Critical Phenomena in Alloys, Magnets, and Superconductors" 
edited by R. E. Mills, E. Ascher and R. I. Jaffee (McGraw-Hill, 1970).

\bibitem{50Nambu}
Y. Nambu,  
Prog. Theor. Phys. 5, 1 (1950).


\bibitem{01Briegel}
H. J. Briegel and R. Raussendorf, 
Phys. Rev. Lett. 86, 910 (2001).

\bibitem{01Raussendorf}
R. Raussendorf and H. J. Briegel, 
Phys. Rev. Lett. 86, 5188 (2001).

\bibitem{03Raussendorf}
R. Raussendorf, D. E. Browne, and H. J. Briegel, 
Phys. Rev. A 68, 022312 (2003).


\bibitem{71Suzuki} 
M. Suzuki, 
Prog. Theor. Phys. 46, 1337 (1971).


\bibitem{04Pachos}
J. K. Pachos and M. B. Plenio, 
Phys. Rev. Lett. 93, 056402 (2004).

\bibitem{05Kopp}
A. Kopp, and S. Chakravarty, 
Nat. Phys.1, 53 (2005).

\bibitem{09Doherty}
A. C. Doherty and S. D. Bartlett, 
Phys. Rev. Lett. 103, 020506 (2009).

\bibitem{09Skrovseth}
S. O. Skr\o vseth and S. D. Bartlett, 
Phys. Rev. A 80, 022316 (2009).

\bibitem{11Smacchia}
P. Smacchia, L. Amico, P. Facchi, R. Fazio, G. Florio, S. Pascazio, and V. Vedral, 
Phys. Rev. A 84, 022304 (2011).

\bibitem{11Son}
W. Son, L. Amico, R. Fazio, A. Hamma, S. Pascazio, and V. Vedral, 
Europhys. Lett. 95, 50001 (2011).

\bibitem{12Montes}
S. Montes and A. Hamma, 
Phys. Rev. E 86, 021101 (2012).

\bibitem{12Son}
W. Son, L. Amico, and V. Vedral,  
Quantum Information Processing 11, 1961 (2012). 

\bibitem{13Cui}
J. Cui, L. Amico, H. Fan, M. Gu, A. Hamma, and V. Vedral, 
Phys. Rev. B 88, 125117 (2013).

\bibitem{14Giampaolo}
S. M. Giampaolo and B. C. Hiesmayr, 
New J. Phys. 16, 093033 (2014). 


\bibitem{15Bridgeman}
J. C. Bridgeman, A. O’Brien, S. D. Bartlett, and A. C.Doherty, 
Phys. Rev. B 91, 165129 (2015).

\bibitem{15Ohta}
T. Ohta, S. Tanaka, I. Danshita, and K. Totsuka, 
J. Phys. Soc. Jpn. 84, 063001 (2015). 

\bibitem{15Lahtinen}
V. Lahtinen and E. Ardonne, 
Phys. Rev. Lett. 115, 237203 (2015).

\bibitem{16Ohta}
T. Ohta, S. Tanaka, I. Danshita, and K. Totsuka, 
Phys. Rev. B 93, 165423 (2016).


\bibitem{07Feng}
X. Y. Feng, G. M. Zhang, and T. Xiang, 
Phys. Rev. Lett. 98, 087204 (2007).


\bibitem{16Minami}
K. Minami, 
J. Phys. Soc. Jpn. 85, 024003 (2016). 


\bibitem{96Gottesman}
D. Gottesman, 
Phys. Rev. A 54, 1862 (1996). 

\bibitem{97Calderbank}
A. R. Calderbank, E. M. Rains, P. W. Shor, and N. J. A. Sloane, 
Phys. Rev. Lett. 78, 405 (1997). 



\bibitem{61Lieb}
E. Lieb, T. Schultz, and D. Mattis,  
Ann. Phys. 16, 407 (1961).


\bibitem{86GehlenRittenbergRuegg}
G. von Gehlen, V. Rittenberg, and H. Ruegg, 
J. Phys. A: Math. Gen. 19, 107 (1986).

\bibitem{86BloteCardyNightingale}
H. W. J. Bl\"{o}te, J. L. Cardy, and M.P. Nightingale, 
Phys. Rev. Lett. 56, 742 (1986). 

\bibitem{86Affleck}
I. Affleck, 
Phys. Rev. Lett. 56, 746 (1986). 

\bibitem{87Henkel}
M. Henkel, 
J. Phys. A: Math. Gen. 20, 995 (1987).



\bibitem{13Mansson}
T. M$\mathring{\rm a}$nsson, V. Lahtinen, J. Suorsa, and E. Ardonne, 
Phys. Rev. B 88, 041403(R) (2013).

\bibitem{14Lahtinen}
V. Lahtinen, T. M$\mathring{\rm a}$nsson, and E. Ardonne, 
Phys. Rev. B 89, 014409 (2014).



\bibitem{04Verstraete}
F. Verstraete, M. Popp, and J. I. Cirac, 
Phys. Rev. Lett. 92, 027901 (2004).

\bibitem{05PoppVerstraete}
M. Popp, F. Verstraete,  M. A. Mart\'{i}n-Delgado, and J. I. Cirac, 
Phys. Rev. A 71, 042306 (2005).


\bibitem{10Gross}
D. Gross and J. Eisert, 
Phys. Rev. A 82, 040303(R) (2010).

\bibitem{12Else}
D. V. Else, I. Schwarz, S. D. Bartlett, and A. C. Doherty, 
Phys. Rev. Lett. 108, 240505 (2012).




\end{thebibliography}



\begin{table}
\caption{\label{table1}
Examples of solvable Hamiltonians 
$-\beta{\cal H}(k, n, l)$ 
obtained as a linear combination of $\eta_{j}$ and their shifted operators. 
Generalizations following (\ref{Hamiltonian}) of these Hamiltonians  can be found in Table 2. 
}
\footnotesize
\begin{tabular}{@{}lll}
\hline
\hline
$(k, n, l)$ & 
$-\beta{\cal H}(k, n, l)$
&
\\
\hline
$(1, n, l)$ & 
$\displaystyle 
K_1\sum_{j=1}^N
\Big(\prod_{\nu=1}^{n}\sigma_{j+\nu-1}^x\Big)
\Big(\prod_{\nu=1}^{l}\sigma_{j+n+\nu-1}^z\Big)
\Big(\prod_{\nu=1}^{n}\sigma_{j+n+l+\nu-1}^x\Big)
$
&
$\displaystyle 
+
K_2\sum_{j=1}^N
\sigma_{j}^x
\sigma_{j+1}^x
$ \\
$(2, n, l)$ & 
$\displaystyle 
K_1\sum_{j=1}^N
\Big(\prod_{\nu=1}^{n}\sigma_{j+\nu-1}^x\Big)
\Big(\prod_{\nu=1}^{l}\sigma_{j+n+\nu-1}^z\Big)
\Big(\prod_{\nu=1}^{n}\sigma_{j+n+l+\nu-1}^x\Big)
$
&
$\displaystyle 
+
K_2\sum_{j=1}^N
\sigma_{j}^y
\sigma_{j+1}^y
$ \\
$(3, n, l)$ & 
$\displaystyle 
K_1\sum_{j=1}^N
\Big(\prod_{\nu=1}^{n}\sigma_{j+\nu-1}^x\Big)
\sigma_{j+n}^z
\Big(\prod_{\nu=1}^{n}\sigma_{j+n+\nu}^x\Big)
$
&
$\displaystyle 
+
K_2\sum_{j=1}^N
\sigma_{j}^x
\Big(\prod_{\nu=1}^{l}{\bf 1}_{j+\nu}\Big)
\sigma_{j+l+1}^x
$ \\
$(4, n, -)$ & 
$\displaystyle 
K_1\sum_{j=1}^N
\sigma_{j}^x
\Big(\prod_{\nu=1}^{n}\sigma_{j+\nu}^z\Big)
\sigma_{j+n+1}^x
$
&
$\displaystyle 
+
K_2\sum_{j=1}^N
\sigma_{j}^z
\Big(\prod_{\nu=1}^{n}{\bf 1}_{j+\nu}\Big)
\sigma_{j+n+1}^z
$ \\
$(5, n, -)$ & 
$\displaystyle 
K_1\sum_{j=1}^N
\Big(\prod_{\nu=1}^{n}\sigma_{j+\nu-1}^z\Big)
$
&
$\displaystyle 
+
K_2\sum_{j=1}^N
\sigma_{j}^x
\sigma_{j+1}^x
$ \\
$(6, n, -)$ & 
$\displaystyle 
K_1\sum_{j=1}^N
\sigma_{j}^x
\Big(\prod_{\nu=1}^{n}{\bf 1}_{j+\nu}\Big)
\sigma_{j+n+1}^x
$
&
$\displaystyle 
+
K_2\sum_{j=1}^N
\sigma_{j}^z
$ \\
$(7, n, l)$ & 
$\displaystyle 
K_1\sum_{j=1}^N
\sigma_{j}^x
\Big(\prod_{\nu=1}^{n}\sigma_{j+\nu}^z\Big)
\sigma_{j+n+1}^x
$
&
$\displaystyle 
+
K_2\sum_{j=1}^N
\sigma_{j}^x
\Big(\prod_{\nu=1}^{l}\sigma_{j+\nu}^z\Big)
\sigma_{j+l+1}^x
$ \\
$(8, n, l)$ & 
$\displaystyle 
K_1\sum_{j=1}^N
\sigma_{j}^x
\Big(\prod_{\nu=1}^{n}\sigma_{j+\nu}^z\Big)
\sigma_{j+n+1}^x
$
&
$\displaystyle 
+
K_2\sum_{j=1}^N
\sigma_{j}^y
\Big(\prod_{\nu=1}^{l}\sigma_{j+\nu}^z\Big)
\sigma_{j+l+1}^y
$ \\
$(9, n, -)$ & 
$\displaystyle 
K_1\sum_{j=1}^N
\sigma_{j}^x\sigma_{j+1}^x\sigma_{j+2}^x
\Big(\prod_{\nu=1}^{n}\sigma_{j+2+\nu}^z\Big)
\sigma_{j+n+3}^x\sigma_{j+n+4}^x\sigma_{j+n+5}^x
$
&
$\displaystyle 
+
K_2\sum_{j=1}^N
\sigma_{j}^x
\Big(\prod_{\nu=1}^{n+2}\sigma_{j+\nu}^z\Big)
\sigma_{j+n+3}^x
$ \\
$(10, n, -)$ & 
$\displaystyle 
K_1\sum_{j=1}^N
\sigma_{j}^x\sigma_{j+1}^x
\Big(\prod_{\nu=1}^{n}\sigma_{j+1+\nu}^z\Big)
\sigma_{j+n+2}^x\sigma_{j+n+3}^x
$
&
$\displaystyle 
+
K_2\sum_{j=1}^N
\Big(\prod_{\nu=1}^{n+2}\sigma_{j+\nu-1}^z\Big)
$ \\
$(11, -, -)$ & 
$\displaystyle 
K_1\sum_{j=1}^N
\sigma_{j}^x\sigma_{j+1}^x
\sigma_{j+2}^z
\sigma_{j+3}^x\sigma_{j+4}^x
$
&
$\displaystyle 
+
K_2\sum_{j=1}^N
\sigma_{j}^x
\sigma_{j+1}^z
\sigma_{j+2}^x
$ \\
\hline
\hline
\end{tabular}\\

\end{table}
\normalsize


\begin{table}
\caption{\label{table2}
Each $(k, n, l)$ provides a solvable Hamiltonian, 
where six interactions in (\ref{Hamiltonian}) are explicitly written. 
One can find the operators $\{\eta^{(k, n, l)}_{j}\}$ 
in the second and third row of the first column, i.e. 
$\eta^{(k, n, l)}_{2j-1}=+2i\varphi^{(k, n, l)}_2(j)\varphi^{(k, n, l)}_1(j)$ and 
$\eta^{(k, n, l)}_{2j}=-2i\varphi^{(k, n, l)}_2(j)\varphi^{(k, n, l)}_1(j+1)$, 
from which a solvable series of interactions are generated. 
The initial operator $\eta^{(k, n, l)}_0$ 
and the transformation $\varphi^{(k, n, l)}_{j}$ that diagonalize the Hamiltonian 
are given in the last row. 
The first case $(k, n, l)=(1, 0, 1)$ 
includes the transverse Ising model, the XY model, and the cluster model, as special cases. 
The second case $(k, n, l)=(2, 1, 0)$ 
includes the one-dimensional Kitaev model. 
The operators $\sigma^x_{j}\sigma^z_{j+1}\sigma^x_{j+2}$ are the stabilizers of the cluster state. 
Models except the cases $(1, 0, 1)$ and $(2, 1, 0)$ 
cannot be solved by the Jordan-Wigner transformation. 
}
\footnotesize
\begin{tabular}{@{}ll}
&
\\
\multicolumn{2}{l}{$(k, n, l)$}
\\
\hline
\hline
$-2i\varphi^{(k, n, l)}_2(j)\varphi^{(k, n, l)}_1(j-1)$ & 
\\
$+2i\varphi^{(k, n, l)}_2(j)\varphi^{(k)}_1(j)\hspace{1.13cm}=\eta^{(k, n, l)}_{2j-1}$ & 
\hspace{0.2cm}
$-2i\varphi^{(k, n, l)}_1(j)\varphi^{(k, n, l)}_1(j+1)$ 
\\
$-2i\varphi^{(k, n, l)}_2(j)\varphi^{(k, n, l)}_1(j+1) \hspace{0.1cm}=\eta^{(k, n, l)}_{2j}$ & 
\hspace{0.2cm}
$-2i\varphi^{(k, n, l)}_2(j)\varphi^{(k, n, l)}_2(j+1)$ 
\\
$+2i\varphi^{(k, n, l)}_2(j)\varphi^{(k, n, l)}_1(j+2)$ & 
\\
\hline
$\eta^{(k, n, l)}_0$ &
$\varphi^{(k, n, l)}_{2j}$ 
\hspace{0.4cm}
and 
\hspace{0.6cm}
$\varphi^{(k, n, l)}_{2j+1}$ 
\hspace{0.8cm}
$(j=0, 1, 2, 3, \ldots)$
\hspace{0.6cm}
\\
\hline
\end{tabular}
\begin{tabular}{@{}ll}
&
\\
&
\\
\multicolumn{2}{l}{$(1, 0, 1)$}
\\
\hline
$\sigma_{j-1}^y\sigma_{j}^y$ & 
\\
$\sigma^z_{j}$ & 
$\sigma_{j}^y\sigma_{j+1}^x$ 
\\
$\sigma_{j}^x\sigma_{j+1}^x$ & 
$\sigma_{j}^x\sigma_{j+1}^y$ 
\\
$\sigma_{j}^x\sigma_{j+1}^z\sigma_{j+2}^x$ & 
\\
\hline
$i\sigma^x_{1}$ &
$\displaystyle \varphi^{(1, 0, 1)}_{2j}=\frac{1}{\sqrt{2}}(\prod_{\nu=1}^{j}\sigma^z_{\nu})\sigma^x_{j+1}$ 
\hspace{0.6cm}
$\displaystyle \varphi^{(1, 0, 1)}_{2j+1}=\frac{1}{\sqrt{2}}(\prod_{\nu=1}^{j}\sigma^z_{\nu})\sigma^y_{j+1}$
\\
\hline
&
\\
\multicolumn{2}{l}{$(2, 1, 0)$}
\\
\hline
$\sigma_{2j-3}^x\sigma_{2j-2}^z\sigma_{2j-1}^z\sigma_{2j}^x$ & 
\\
$\sigma^x_{2j-1}\sigma^x_{2j}$ & 
$\sigma_{2j-1}^x\sigma_{2j}^z\sigma_{2j+1}^y$ 
\\
$\sigma^y_{2j}\sigma^y_{2j+1}$ & 
$\sigma_{2j}^y\sigma_{2j+1}^z\sigma_{2j+2}^x$ 
\\
$\sigma_{2j}^y\sigma_{2j+1}^z\sigma_{2j+2}^z\sigma_{2j+3}^y$ & 
\\
\hline
$i\sigma^y_{1}$ &
$\displaystyle \varphi^{(2, 1, 0)}_{2j}=\frac{(-1)^j}{\sqrt{2}}(\prod_{\nu=1}^{2j}\sigma^z_{\nu})\sigma^y_{2j+1}$ 
\hspace{0.6cm}
$\displaystyle \varphi^{(2, 1, 0)}_{2j+1}=\frac{(-1)^j}{\sqrt{2}}(\prod_{\nu=1}^{2j+1}\sigma^z_{\nu})\sigma^x_{2j+2}$
\\
\hline
\end{tabular}
\end{table}

\newpage
\begin{table}
\footnotesize
\begin{tabular}{@{}ll}
&
\\
\multicolumn{2}{l}{$(3, 1, 1)$}
\\
\hline
$\sigma_{2j-3}^x\sigma_{2j-2}^y 1_{2j-1}\sigma_{2j}^y\sigma_{2j+1}^x$ & 
\\
$\sigma_{2j-1}^x\sigma_{2j}^z\sigma_{2j+1}^x$ & 
$\sigma_{2j-1}^x\sigma_{2j}^y\sigma_{2j+1}^x\sigma_{2j+2}^x$ 
\\
$\sigma_{2j}^x 1_{2j+1}\sigma_{2j+2}^x$ & 
$\sigma_{2j}^x\sigma_{2j+1}^x\sigma_{2j+2}^y\sigma_{2j+3}^x$ 
\\
$\sigma_{2j}^x\sigma_{2j+1}^x\sigma_{2j+2}^z\sigma_{2j+3}^x\sigma_{2j+4}^x$ & 
\\
\hline
$i\sigma^x_{1}\sigma^x_{2}$ &
$\displaystyle \varphi^{(3, 1, 1)}_{2j}=\frac{1}{\sqrt{2}}(\prod_{\nu=1}^{j}1_{2\nu-1}\sigma^z_{2\nu})\sigma^x_{2j+1}\sigma^x_{2j+2}$ 
\\
 &
$\displaystyle \varphi^{(3, 1, 1)}_{2j+1}=\frac{1}{\sqrt{2}}(\prod_{\nu=1}^{j}1_{2\nu-1}\sigma^z_{2\nu})1_{2j+1}\sigma^y_{2j+2}\sigma^x_{2j+3}$
\\
\hline
&
\\
\multicolumn{2}{l}{$(3, 1, 2)$}
\\
\hline
$\sigma_{3j-5}^x\sigma_{3j-4}^y\sigma_{3j-3}^x\sigma_{3j-2}^x\sigma_{3j-1}^y\sigma_{3j}^x$ & 
\\
$\sigma_{3j-2}^x\sigma_{3j-1}^z\sigma_{3j}^x$ & 
$\sigma_{3j-2}^x\sigma_{3j-1}^y\sigma_{3j}^x 1_{3j+1}\sigma_{3j+2}^x$ 
\\
$\sigma_{3j-1}^x 1_{3j} 1_{3j+1}\sigma_{3j+2}^x$ & 
$\sigma_{3j-1}^x 1_{3j}\sigma_{3j+1}^x\sigma_{3j+2}^y\sigma_{3j+3}^x$ 
\\
$\sigma_{3j-1}^x 1_{3j}\sigma_{3j+1}^x\sigma_{3j+2}^z\sigma_{3j+3}^x 1_{3j+4}\sigma_{3j+5}^x$ & 
\\
\hline
$i1_{1}\sigma^x_{2}1_{3}$ &
$\displaystyle \varphi^{(3, 1, 2)}_{2j}=\frac{1}{\sqrt{2}}(\prod_{\nu=1}^{j}\sigma^x_{3\nu-2}\sigma^z_{3\nu-1}\sigma^x_{3\nu})1_{3j+1}\sigma^x_{3j+2}$ 
\\
 &
$\displaystyle \varphi^{(3, 1, 2)}_{2j+1}=\frac{1}{\sqrt{2}}(\prod_{\nu=1}^{j}\sigma^x_{3\nu-2}\sigma^z_{3\nu-1}\sigma^x_{3\nu})\sigma^x_{3j+1}\sigma^y_{3j+2}\sigma^x_{3j+3}$
\\
\hline
\end{tabular}
\begin{tabular}{@{}ll}
&
\\
\multicolumn{2}{l}{$(3, 1, 3)$}
\\
\hline
$\sigma_{4j-7}^x\sigma_{4j-6}^y\sigma_{4j-5}^x 1_{4j-4}\sigma_{4j-3}^x\sigma_{4j-2}^y\sigma_{4j-1}^x$ & 
\\
$\sigma_{4j-3}^x\sigma_{4j-2}^z\sigma_{4j-1}^x$ & 
$\sigma_{4j-3}^x\sigma_{4j-2}^y\sigma_{4j-1}^x 1_{4j} 1_{4j+1}\sigma_{4j+2}^x$ 
\\
$\sigma_{4j-2}^x 1_{4j-1}1_{4j}1_{4j+1}\sigma_{4j+2}^x$ & 
$\sigma_{4j-2}^x 1_{4j-1} 1_{4j}\sigma_{4j+1}^x\sigma_{4j+2}^y\sigma_{4j+3}^x$ 
\\
$\sigma_{4j-2}^x 1_{4j-1} 1_{4j}\sigma_{4j+1}^x\sigma_{4j+2}^z\sigma_{4j+3}^x 1_{4j+4} 1_{4j+5}\sigma_{4j+6}^x$ & 
\\
\hline
$i1_{1}\sigma^x_{2}$ &
$\displaystyle \varphi^{(3, 1, 3)}_{2j}=\frac{1}{\sqrt{2}}(\prod_{\nu=1}^{j}\sigma^x_{4\nu-3}\sigma^z_{4\nu-2}\sigma^x_{4\nu-1}1_{4\nu})1_{4j+1}\sigma^x_{4j+2}$ 
\\
&
$\displaystyle \varphi^{(3, 1, 3)}_{2j+1}=\frac{1}{\sqrt{2}}(\prod_{\nu=1}^{j}\sigma^x_{4\nu-3}\sigma^z_{4\nu-2}\sigma^x_{4\nu-1}1_{4\nu})\sigma^x_{4j+1}\sigma^y_{4j+2}\sigma^x_{4j+3}$
\\
\hline
\end{tabular}
\begin{tabular}{@{}ll}
&
\\
\multicolumn{2}{l}{$(4, 1, -)$}
\\
\hline
$\sigma_{4j-7}^x\sigma_{4j-6}^z \sigma_{4j-5}^y 1_{4j-4}\sigma_{4j-3}^y\sigma_{4j-2}^z\sigma_{4j-1}^x$ & 
\\
$\sigma_{4j-3}^x\sigma_{4j-2}^z\sigma_{4j-1}^x$ & 
$\sigma_{4j-3}^x\sigma_{4j-2}^z\sigma_{4j-1}^y 1_{4j}\sigma_{4j+1}^z$ 
\\
$\sigma_{4j-1}^z 1_{4j}\sigma_{4j+1}^z$ & 
$\sigma_{4j-1}^z 1_{4j}\sigma_{4j+1}^y\sigma_{4j+2}^z\sigma_{4j+3}^x$ 
\\
$\sigma_{4j-1}^z 1_{4j}\sigma_{4j+1}^y\sigma_{4j+2}^z\sigma_{4j+3}^y 1_{4j+4}\sigma_{4j+5}^z$ & 
\\
\hline
$-i\sigma^z_{1}$ 
&
$\displaystyle \varphi^{(4, 1, -)}_{2j}=\frac{(-1)^{j-1}}{\sqrt{2}}(\prod_{\nu=1}^{j}\sigma^y_{4\nu-3}\sigma^z_{4\nu-2}\sigma^y_{4\nu-1}1_{4\nu})\sigma^z_{4j+1}$ 
\\
&
$\displaystyle \varphi^{(4, 1, -)}_{2j+1}=\frac{(-1)^{j}}{\sqrt{2}}(\prod_{\nu=1}^{j}\sigma^y_{4\nu-3}\sigma^z_{4\nu-2}\sigma^y_{4\nu-1}1_{4\nu})\sigma^y_{4j+1}\sigma^z_{4j+2}\sigma^x_{4j+3}$ 
\\
\hline
\end{tabular}
\end{table}

\newpage
\begin{table}
\footnotesize
\begin{tabular}{@{}ll}
&
\\
\multicolumn{2}{l}{$(4, 2, -)$}
\\
\hline
$\sigma_{6j-11}^x\sigma_{6j-10}^z\sigma_{6j-9}^z\sigma_{6j-8}^y 1_{6j-7} 1_{6j-6}\sigma_{6j-5}^y\sigma_{6j-4}^z\sigma_{6j-3}^z\sigma_{6j-2}^x$ & 
\\
$\sigma_{6j-5}^x\sigma_{6j-4}^z\sigma_{6j-3}^z\sigma_{6j-2}^x$ & 
$\sigma_{6j-5}^x\sigma_{6j-4}^z\sigma_{6j-3}^z\sigma_{6j-2}^y 1_{6j-1} 1_{6j}\sigma_{6j+1}^z$
\\
$\sigma_{6j-2}^z\ 1_{6j-1} 1_{6j}\sigma_{6j+1}^z$ & 
$\sigma_{6j-2}^z 1_{6j-1} 1_{6j}\sigma_{6j+1}^y\sigma_{6j+2}^z\sigma_{6j+3}^z\sigma_{6j+4}^x$
\\
$\sigma_{6j-2}^z 1_{6j-1j} 1_{6j}\sigma_{6j+1}^y\sigma_{6j+2}^z\sigma_{6j+3}^z\sigma_{6j+4}^y 1_{6j+5} 1_{6j+6}\sigma_{6j+7}^z$ & 
\\
\hline
\end{tabular}
\begin{tabular}{@{}ll}
$-i\sigma^z_{1}$ 
&
\hspace{3.2cm}
$\displaystyle \varphi^{(4, 2, -)}_{2j}=\frac{(-1)^{j-1}}{\sqrt{2}}(\prod_{\nu=1}^{j}\sigma^y_{6\nu-5}\sigma^z_{6\nu-4}\sigma^z_{6\nu-3}\sigma^y_{6\nu-2}1_{6\nu-1}1_{6\nu})\sigma^z_{6j+1}$
\\
&
\hspace{3.2cm}
$\displaystyle \varphi^{(4, 2, -)}_{2j+1}=\frac{(-1)^{j}}{\sqrt{2}}(\prod_{\nu=1}^{j}\sigma^y_{6\nu-5}\sigma^z_{6\nu-4}\sigma^z_{6\nu-3}\sigma^y_{6\nu-2}1_{6\nu-1}1_{6\nu})\sigma^y_{6j+1}\sigma^z_{6j+2}\sigma^z_{6j+3}\sigma^x_{6j+4}$
\\
\hline
\end{tabular}
%
%
\begin{tabular}{@{}ll}
&
\\
\multicolumn{2}{l}{$(5, 3, -)$}
\\
\hline
$\sigma_{3j-5}^z\sigma_{3j-4}^z \sigma_{3j-3}^y\sigma_{3j-2}^y\sigma_{3j-1}^z\sigma_{3j}^z$ & 
\\
$\sigma_{3j-2}^z\sigma_{3j-1}^z\sigma_{3j}^z$ & 
$\sigma_{3j-2}^z\sigma_{3j-1}^z\sigma_{3j}^y\sigma_{3j+1}^x$ 
\\
$\sigma_{3j}^x\sigma_{3j+1}^x$ & 
$\sigma_{3j}^x\sigma_{3j+1}^y\sigma_{3j+2}^z\sigma_{3j+3}^z$ 
\\
$\sigma_{3j}^x\sigma_{3j+1}^y\sigma_{3j+2}^z\sigma_{3j+3}^y\sigma_{3j+4}^x$ & 
\\
\hline
$i\sigma^x_{1}$ &
$\displaystyle \varphi^{(5, 3, -)}_{2j}=\frac{(-1)^j}{\sqrt{2}}(\prod_{\nu=1}^{j}\sigma^y_{3\nu-2}\sigma^z_{3\nu-1}\sigma^y_{3\nu})\sigma^x_{3j+1}$ 
\\
 &
$\displaystyle \varphi^{(5, 3, -)}_{2j+1}=\frac{(-1)^j}{\sqrt{2}}(\prod_{\nu=1}^{j}\sigma^y_{3\nu-2}\sigma^z_{3\nu-1}\sigma^y_{3\nu})\sigma^y_{3j+1}\sigma^z_{3j+2}\sigma^z_{3j+3}$
\\
\hline
&
\\
\multicolumn{2}{l}{$(1, 2, 1)$}
\\
\hline
$\sigma_{j-1}^x 1_{j}\sigma_{j+1}^z\sigma_{j+2}^z 1_{j+3}\sigma_{j+4}^x$ & 
\\
$\sigma_{j}^x\sigma_{j+1}^x\sigma_{j+2}^z\sigma_{j+3}^x\sigma_{j+4}^x$ & 
$\sigma_{j}^x\sigma_{j+1}^x\sigma_{j+2}^y 1_{j+3}\sigma_{j+4}^x$
\\
$\sigma_{j+2}^x\sigma_{j+3}^x$ & 
$\sigma_{j+2}^x 1_{j+3}\sigma_{j+4}^y\sigma_{j+5}^x\sigma_{j+6}^x$
\\
$\sigma_{j+2}^x 1_{j+3}\sigma_{j+4}^z 1_{j+5}\sigma_{j+6}^x$ & 
\\
\hline
$i\sigma^y_{1}\sigma^z_{2}\sigma^x_{3}\sigma^x_{4}$ &
$\displaystyle \varphi^{(1, 2, 1)}_{2j}=\frac{1}{\sqrt{2}}(\prod_{\nu=1}^{j}\sigma^z_{\nu})\sigma^y_{j+1}\sigma^z_{j+2}\sigma^x_{j+3}\sigma^x_{j+4}$ 
\\
&
$\displaystyle \varphi^{(1, 2, 1)}_{2j+1}=\frac{1}{\sqrt{2}}(\prod_{\nu=1}^{j}\sigma^z_{\nu})\sigma^z_{j+1}\sigma^y_{j+2}\sigma^y_{j+3} 1_{j+4}\sigma^x_{j+5}$
\\
\hline
\end{tabular}
\begin{tabular}{@{}ll}
&
\\
\multicolumn{2}{l}{$(2, 2, 1)$}
\\
\hline
$\sigma_{5j-9}^x\sigma_{5j-8}^x\sigma_{5j-7}^z\sigma_{5j-6}^x\sigma_{5j-5}^z\sigma_{5j-4}^z\sigma_{5j-3}^x\sigma_{5j-2}^z\sigma_{5j-1}^x\sigma_{5j}^x$ & 
\\
$\sigma_{5j-4}^x\sigma_{5j-3}^x\sigma_{5j-2}^z\sigma_{5j-1}^x\sigma_{5j}^x$ & 
$\sigma_{5j-4}^x\sigma_{5j-3}^x\sigma_{5j-2}^z\sigma_{5j-1}^x\sigma_{5j}^z\sigma_{5j+1}^y$ 
\\
$\sigma_{5j}^y\sigma_{5j+1}^y$ & 
$\sigma_{5j}^y\sigma_{5j+1}^z\sigma_{5j+2}^x\sigma_{5j+3}^z\sigma_{5j+4}^x\sigma_{5j+5}^x$ 
\\
$\sigma_{5j}^y\sigma_{5j+1}^z\sigma_{5j+2}^x\sigma_{5j+3}^z\sigma_{5j+4}^x\sigma_{5j+5}^z\sigma_{5j+6}^y$ & 
\\
\hline
\end{tabular}
\begin{tabular}{@{}ll}
$i\sigma^y_{1}$ &
\hspace{1.6cm}
$\displaystyle \varphi^{(2, 2, 1)}_{2j}=\frac{(-1)^j}{\sqrt{2}}(\prod_{\nu=1}^{j}\sigma^z_{5\nu-4}\sigma^x_{5\nu-3}\sigma^z_{5\nu-2}\sigma^x_{5\nu-1}\sigma^z_{5\nu})\sigma^y_{5j+1}$ 
\\
 &
\hspace{1.6cm}
$\displaystyle \varphi^{(2, 2, 1)}_{2j+1}=\frac{(-1)^j}{\sqrt{2}}(\prod_{\nu=1}^{j}\sigma^z_{5\nu-4}\sigma^x_{5\nu-3}\sigma^z_{5\nu-2}\sigma^x_{5\nu-1}\sigma^z_{5\nu})\sigma^z_{5j+1}\sigma^x_{5j+2}\sigma^z_{5j+3} \sigma^x_{5j+4}\sigma^x_{5j+5}$
\\
\hline
\end{tabular}
\end{table}
\normalsize
\begin{table}
\footnotesize
\begin{tabular}{@{}ll}
&
\\
\multicolumn{2}{l}{$(4, 2, 2)$}
\\
\hline
$\sigma_{3j-5}^x\sigma_{3j-4}^x\sigma_{3j-3}^y 1_{3j-2} 1_{3j-1}\sigma_{3j}^y\sigma_{3j+1}^x\sigma_{3j+2}^x$ & 
\\
$\sigma_{3j-2}^x\sigma_{3j-1}^x\sigma_{3j}^z\sigma_{3j+1}^x\sigma_{3j+2}^x$ & 
$\sigma_{3j-2}^x\sigma_{3j-1}^x\sigma_{3j}^y\sigma_{3j+1}^x\sigma_{3j+2}^x\sigma_{3j+3}^x$ 
\\
$\sigma_{3j}^x 1_{3j+1} 1_{3j+2}\sigma_{3j+3}^x$ & 
$\sigma_{3j}^x\sigma_{3j+1}^x\sigma_{3j+2}^x\sigma_{3j+3}^y\sigma_{3j+4}^x\sigma_{3j+5}^x$ 
\\
$\sigma_{3j}^x\sigma_{3j+1}^x\sigma_{3j+2}^x\sigma_{3j+3}^z\sigma_{3j+4}^x\sigma_{3j+5}^x\sigma_{3j+6}^x$ & 
\\
\hline
$i\sigma_{1}^x\sigma_{2}^x\sigma_{3}^x$ &
$\displaystyle \varphi^{(4, 2, 2)}_{2j}=\frac{1}{\sqrt{2}}(\prod_{\nu=1}^{j}1_{3\nu-2}1_{3\nu-1}\sigma^z_{3\nu})\sigma_{3j+1}^x\sigma_{3j+2}^x\sigma_{3j+3}^x$ 
\\
 &
$\displaystyle \varphi^{(4, 2, 2)}_{2j+1}=\frac{1}{\sqrt{2}}(\prod_{\nu=1}^{j}1_{3\nu-2}1_{3\nu-1}\sigma^z_{3\nu})1_{3j+1}1_{3j+2}\sigma_{3j+3}^y\sigma_{3j+4}^x\sigma_{3j+5}^x$
\\
\hline
\end{tabular}
\begin{tabular}{@{}ll}
&
\\
\multicolumn{2}{l}{$(11, -, -)$}
\\
\hline
$\sigma_{4j-7}^x\sigma_{4j-6}^x\sigma_{4j-5}^z 1_{4j-4}\sigma_{4j-3}^z 1_{4j-2}\sigma_{4j-1}^z\sigma_{4j}^x\sigma_{4j+1}^x$ & 
\\
$\sigma_{4j-3}^x\sigma_{4j-2}^x\sigma_{4j-1}^z\sigma_{4j}^x\sigma_{4j+1}^x$ & 
$\sigma_{4j-3}^x\sigma_{4j-2}^x\sigma_{4j-1}^z 1_{4j}\sigma_{4j+1}^y\sigma_{4j+2}^x$ 
\\
$\sigma_{4j}^x\sigma_{4j+1}^z\sigma_{4j+2}^x$& 
$\sigma_{4j}^x\sigma_{4j+1}^y 1_{4j+2}\sigma_{4j+3}^z\sigma_{4j+4}^x\sigma_{4j+5}^x$ 
\\
$\sigma_{4j}^x\sigma_{4j+1}^y 1_{4j+2}\sigma_{4j+3}^z 1_{4j+4}\sigma_{4j+5}^y\sigma_{4j+6}^x$ & 
\\
\hline
\end{tabular}
\begin{tabular}{@{}ll}
$i\sigma_{1}^y\sigma_{2}^x$ &
\hspace{1.6cm}
$\displaystyle \varphi^{(11, -, -)}_{2j}=\frac{1}{\sqrt{2}}(\prod_{\nu=1}^{j}\sigma^z_{4\nu-3}1_{4\nu-2}\sigma^z_{4\nu-1}1_{4\nu})\sigma_{4j+1}^y\sigma_{4j+2}^x$ 
\\
 &
\hspace{1.6cm}
$\displaystyle \varphi^{(11, -, -)}_{2j}=\frac{1}{\sqrt{2}}(\prod_{\nu=1}^{j}\sigma^z_{4\nu-3}1_{4\nu-2}\sigma^z_{4\nu-1}1_{4\nu})\sigma_{4j+1}^z 1_{4j+2}\sigma_{4j+3}^z\sigma_{4j+4}^x\sigma_{4j+5}^x$ 
\\
\hline
\hline
\end{tabular}
\vspace{12.0cm}

\end{table}

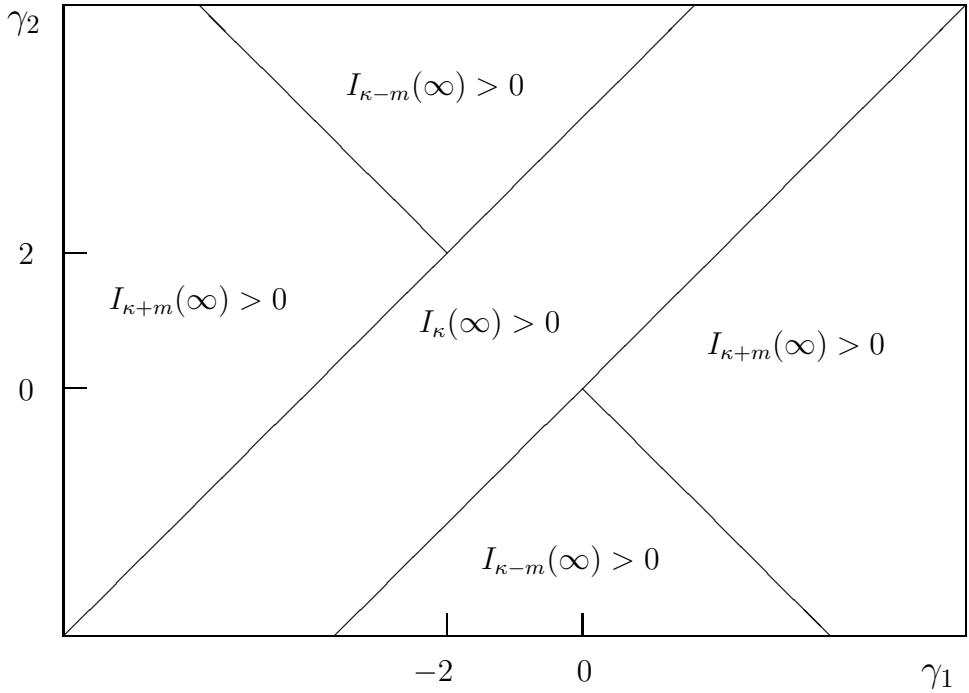
\begin{figure}[b]
\unitlength=1.5mm
\begin{picture}(88, 60)
%
\put(5,4){\line(1,0){80}}
\put(5,60){\line(1,0){80}}
\put(5,4){\line(0,1){56}}
\put(85,4){\line(0,1){56}}
\put(0,58){\mbox{{\large$\gamma_2$}}}
\put(81,0){\mbox{{\large$\gamma_1$}}}
\put(39,4){\line(0,1){2}}
\put(51,4){\line(0,1){2}}
\put(5,26){\line(1,0){2}}
\put(5,38){\line(1,0){2}}
\put(36,0){\mbox{$-2$}}
\put(50.5,0){\mbox{$0$}}
\put(1,25){\mbox{$0$}}
\put(1,37){\mbox{$2$}}
%
\put(5,4){\line(1,1){56}}
\put(29,4){\line(1,1){56}}
\put(51,26){\line(1,-1){22}}
\put(39,38){\line(-1,1){22}}
\put(36.5,31){\mbox{$I_\kappa(\infty)>0$}}
\put(9,33){\mbox{$I_{\kappa+m}(\infty)>0$}}
\put(62,29){\mbox{$I_{\kappa+m}(\infty)>0$}}
\put(30,52){\mbox{$I_{\kappa-m}(\infty)>0$}}
\put(42,10){\mbox{$I_{\kappa-m}(\infty)>0$}}
\end{picture}
\caption{\label{fig:epsart} 
Phase diagram of the model (\ref{HamgenXY}). 
The exponent in (\ref{I_infty}) equals $m/4$ on the critical line 
$\gamma_2=\gamma_1\neq 0$ and 
$\gamma_2=\gamma_1+4\neq 2$, 
and equals $2m/4$ on the critical line 
$\gamma_2=-\gamma_1$ $(\gamma_1<-2, 0<\gamma_1)$. 
$I_\kappa(n)$ is discontinuous at the points $(0, 0)$ and $(-2, 2)$, 
in the limit $n\to\infty$. . 
}
\end{figure}

\begin{table}
\caption{\label{table3}
Stabilizers and the corresponding string order parameters.}
\footnotesize
\begin{tabular}{@{}ll}
\hline
\hline
stabilizers
&
string order parameters (with finite $n$)
\\
\hline
$\displaystyle \sigma^z_{j}$
& 
$
\displaystyle \langle \prod_{\nu=j}^{j+n}\sigma^z_{\nu}\rangle
$
\\
$\displaystyle \sigma^x_{j}\sigma^x_{j+1}$
& 
$
\displaystyle \langle \prod_{\nu=j}^{j+n}\sigma^x_{\nu}\sigma^x_{\nu+1}\rangle
=
\displaystyle \langle \sigma^x_{j}(\prod_{\nu=j+1}^{j+n}1_{\nu})\sigma^x_{j+n+1}\rangle
$
\\
$\displaystyle \sigma^z_{j}\sigma^z_{j+1}\sigma^z_{j+2}$
& 
$
\displaystyle \langle \prod_{\nu=j}^{j+n}\sigma^z_{\nu}\sigma^z_{\nu+1}\sigma^z_{\nu+2}\rangle
=
\displaystyle \langle \sigma^z_{j}1_{j+1}(\prod_{\nu=j+2}^{j+n+1}\sigma^z_{\nu})1_{j+n+2}\sigma^z_{j+n+3}\rangle
$
\\
$\displaystyle \sigma^y_{2j}\sigma^y_{2j+1}$
& 
$
\displaystyle \langle \prod_{\nu=j}^{j+n}\sigma^y_{2\nu}\sigma^y_{2\nu+1}\rangle
=
\displaystyle \langle \prod_{\nu=2j}^{2j+2n+1}\sigma^y_{\nu}\rangle
$
\\
$\displaystyle \sigma^x_{j}\sigma^z_{j+1}\sigma^x_{j+2}$
& 
$
\displaystyle \langle \prod_{\nu=j}^{j+n}\sigma^x_{\nu}\sigma^z_{\nu+1}\sigma^x_{\nu+2}\rangle
=
\displaystyle \langle \sigma^x_{j}\sigma^y_{j+1}(\prod_{\nu=j+2}^{j+n}\sigma^z_{\nu})\sigma^y_{j+n+1}\sigma^x_{j+n+2}\rangle
$
\\
$\displaystyle \sigma^x_{j}\sigma^z_{j+1}\sigma^z_{j+2}\sigma^x_{j+3}$
& 
$
\displaystyle \langle \prod_{\nu=j}^{j+n}\sigma^x_{\nu}\sigma^z_{\nu+1}\sigma^z_{\nu+2}\sigma^x_{\nu+3}\rangle
=
\displaystyle \langle \sigma^x_{j}\sigma^y_{j+1}\sigma^x_{j+2}(\prod_{\nu=j+3}^{j+n}1_{\nu})\sigma^x_{j+n+1}\sigma^y_{j+n+2}\sigma^x_{j+n+3}\rangle
$
\\
$\displaystyle \sigma^x_{j}\sigma^x_{j+1}\sigma^z_{j+2}\sigma^x_{j+3}\sigma^x_{j+4}$
& 
$
\displaystyle \langle \prod_{\nu=j}^{j+n}\sigma^x_{\nu}\sigma^z_{\nu+1}\sigma^x_{\nu+2}\rangle
=
\displaystyle \langle \sigma^x_{j} 1_{j+1}\sigma^z_{j+2}\sigma^y_{j+3}(\prod_{\nu=j+4}^{j+n}\sigma^z_{\nu})\sigma^y_{j+n+1}\sigma^z_{j+n+2}1_{j+n+3}\sigma^x_{j+n+4}\rangle
$
\\
$\displaystyle \sigma^x_{j} 1_{j+1}\sigma^x_{j+2}$
& 
$
\displaystyle \langle \prod_{\nu=j}^{j+n}\sigma^x_{\nu} 1_{\nu+1}\sigma^x_{\nu+2}\rangle
=
\displaystyle \langle \sigma^x_{j}\sigma^x_{j+1}(\prod_{\nu=j+2}^{j+n}1_{\nu})\sigma^x_{j+n+1}\sigma^x_{j+n+2}\rangle
$
\\
$\displaystyle \sigma^x_{j} 1_{j+1}1_{j+2}\sigma^x_{j+3}$
& 
$
\displaystyle \langle \prod_{\nu=j}^{j+n}\sigma^x_{\nu} 1_{\nu+1}1_{\nu+2}\sigma^x_{\nu+3}\rangle
=
\displaystyle \langle \sigma^x_{j}\sigma^x_{j+1}\sigma^x_{j+2}(\prod_{\nu=j+3}^{j+n}1_{\nu})\sigma^x_{j+n+1}\sigma^x_{j+n+2}\sigma^x_{j+n+3}\rangle
$
\\
$\displaystyle \sigma^x_{j} 1_{j+1}1_{j+2}1_{j+3}\sigma^x_{j+4}$
& 
$
\displaystyle \langle \prod_{\nu=j}^{j+n}\sigma^x_{\nu} 1_{\nu+1}1_{\nu+2}1_{\nu+3}\sigma^x_{\nu+4}\rangle
=
\displaystyle \langle \sigma^x_{j}\sigma^x_{j+1}\sigma^x_{j+2}\sigma^x_{j+3}(\prod_{\nu=j+4}^{j+n}1_{\nu})\sigma^x_{j+n+1}\sigma^x_{j+n+2}\sigma^x_{j+n+3}\sigma^x_{j+n+4}\rangle
$
\\
$\displaystyle \sigma^z_{j} 1_{j+1}\sigma^z_{j+2}$
& 
$
\displaystyle \langle \prod_{\nu=j}^{j+n}\sigma^z_{\nu} 1_{\nu+1}\sigma^z_{\nu+2}\rangle
=
\displaystyle \langle \sigma^z_{j}\sigma^z_{j+1}(\prod_{\nu=j+2}^{j+n}1_{\nu})\sigma^z_{j+n+1}\sigma^z_{j+n+2}\rangle
$
\\
$\displaystyle \sigma^z_{j} 1_{j+1}1_{j+2}\sigma^z_{j+3}$
& 
$
\displaystyle \langle \prod_{\nu=j}^{j+n}\sigma^z_{\nu} 1_{\nu+1}1_{\nu+2}\sigma^z_{\nu+3}\rangle
=
\displaystyle \langle \sigma^z_{j}\sigma^z_{j+1}\sigma^z_{j+2}(\prod_{\nu=j+3}^{j+n}1_{\nu})\sigma^z_{j+n+1}\sigma^z_{j+n+2}\sigma^z_{j+n+3}\rangle
$
\\
\hline
\hline
\end{tabular}

\end{table}
\normalsize

\end{document}